\documentclass[acmtog,anonymous=false,review=false]{acmart}
\acmSubmissionID{498}

\usepackage{booktabs} % For formal tables

\usepackage{geometrycollective}
\usepackage{BoundaryValueCaching}
\usepackage{mathdefs}

\setcopyright{rightsretained}
\acmJournal{TOG}
\acmYear{2023}
\acmVolume{42}
\acmNumber{4}
\acmArticle{1}
\acmMonth{8}
\acmPrice{15.00}
\acmDOI{10.1145/3592400}

\begin{teaserfigure}
  \centering
  \includegraphics{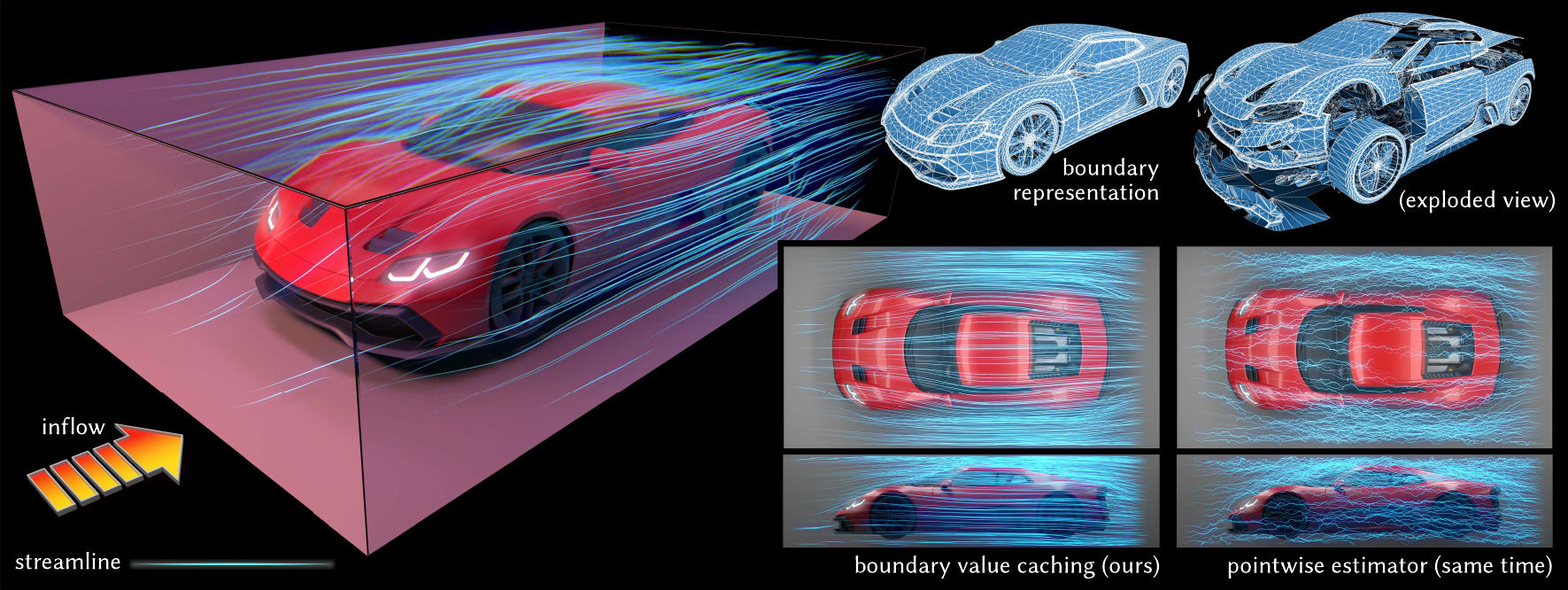}
   \caption{Our caching scheme dramatically reduces the total number of random walks needed to solve partial differential equations relative to classic pointwise Monte Carlo estimators.  Here we show streamlines of a flow in a simulated wind tunnel, computed directly from a low-quality surface mesh originally intended for visualization rather than simulation. \label{fig:teaser}}
\end{teaserfigure}

\begin{document}
% Title portion
\title{Boundary Value Caching for Walk on Spheres}

% DO NOT ENTER AUTHOR INFORMATION FOR ANONYMOUS TECHNICAL PAPER SUBMISSIONS
\author{Bailey Miller}
\authornote{ and $^\dagger$ indicate equal contribution.}
\email{bmmiller@andrew.cmu.edu}
\affiliation{%
  \institution{Carnegie Mellon University}
  %\streetaddress{5000 Forbes Ave}
  %\city{Pittsburgh}
  %\state{PA}
  %\postcode{15213}
  \country{USA}
}
\orcid{0009-0009-0881-0351}

\author{Rohan Sawhney}
\authornotemark[1]
\email{rohansawhney@cs.cmu.edu}
\affiliation{%
  \institution{Carnegie Mellon University and NVIDIA}
  %\streetaddress{5000 Forbes Ave}
  %\city{Pittsburgh}
  %\state{PA}
  %\postcode{15213}
  \country{USA}
}
\orcid{0000-0002-3661-1554}

\author{Keenan Crane}
\authornotemark[2]
\email{kmcrane@cs.cmu.edu}
\affiliation{%
  \institution{Carnegie Mellon University}
  %\streetaddress{5000 Forbes Ave}
  %\city{Pittsburgh}
  %\state{PA}
  %\postcode{15213}
  \country{USA}
}
\orcid{0000-0003-2772-7034}

\author{Ioannis Gkioulekas}
\authornotemark[2]
\email{igkioule@cs.cmu.edu}
\affiliation{%
  \institution{Carnegie Mellon University}
  \streetaddress{5000 Forbes Ave}
  \city{Pittsburgh}
  \state{PA}
  \postcode{15213}
  \country{USA}
}
\orcid{0000-0001-6932-4642}

\begin{abstract}
   Grid-free Monte Carlo methods such as \emph{walk on spheres} can be used to solve elliptic partial differential equations without mesh generation or global solves.  However, such methods independently estimate the solution at every point, and hence do not take advantage of the high spatial regularity of solutions to elliptic problems.  We propose a fast caching strategy which first estimates solution values and derivatives at randomly sampled points along the boundary of the domain (or a local region of interest).  These cached values then provide cheap, output-sensitive evaluation of the solution (or its gradient) at interior points, via a boundary integral formulation.  Unlike classic boundary integral methods, our caching scheme introduces zero statistical bias and does not require a dense global solve.  Moreover we can handle imperfect geometry (e.g., with self-intersections) and detailed boundary/source terms without repairing or resampling the boundary representation.  Overall, our scheme is similar in spirit to \emph{virtual point light} methods from photorealistic rendering: it suppresses the typical salt-and-pepper noise characteristic of independent Monte Carlo estimates, while still retaining the many advantages of Monte Carlo solvers: progressive evaluation, trivial parallelization, geometric robustness, \etc{}\  We validate our approach using test problems from visual and geometric computing.
\end{abstract}





% DO NOT INCLUDE ACKNOWLEDGMENTS IN AN ANONYMOUS SUBMISSION
%\begin{acks}
%
%The authors would like to thank Dr. Maura Turolla of Telecom
%Italia for providing specifications about the application scenario.
%
%The work is supported by the \grantsponsor{GS501100001809}{National
%  Natural Science Foundation of
%  China}{http://dx.doi.org/10.13039/501100001809} under Grant
%No.:~\grantnum{GS501100001809}{61273304\_a}
%and~\grantnum[http://www.nnsf.cn/youngscientists]{GS501100001809}{Young
%  Scientists' Support Program}.
%
%
%\end{acks}

\maketitle

\section{Introduction}
\label{sec:Introduction}

The \emph{walk on spheres (WoS)} method solves problems like the Laplace or Poisson equation by aggregating information from repeated random walks \citep{Muller:1956:WOS,Sawhney:2020:MCGP}.  Like Monte Carlo ray tracing---and unlike conventional partial differential equation (PDE) solvers---it does not require a mesh of the problem domain, nor even a high-quality mesh of its boundary.  This fact makes WoS valuable for problems in visual and geometric computing, as one can directly use imperfect assets from design or visualization to perform simulation and analysis (\figref{teaser}).  However, classic WoS methods estimate the PDE solution \emph{pointwise} and do not share information between sample points, resulting in highly redundant computation.

We propose a simple \emph{boundary value caching (BVC)} scheme, well-suited for problems like visualization, where the solution must be evaluated densely in space. This scheme is enabled by the recent \emph{walk on stars (WoSt)} method \citep{Sawhney:2023:WalkOnStars}, which extends WoS to problems with mixed Neumann and Dirichlet boundary conditions. In particular, we consider PDEs of the form
\begin{equation}
   \label{eq:MainPDE}
   \arraycolsep=0.25em
   \begin{array}{rccl}
      \Delta u - \sigma u &=& f & \quad\text{on}\ \ \Omega \\
      u &=& g & \quad\text{on}\ \ \partial\Omega_D \\
      \frac{\partial u}{\partial n} &=& h & \quad\text{on}\ \ \partial\Omega_N \\
   \end{array}
\end{equation}
where the boundary of the domain \(\Omega \subset \mathbb{R}^n\) is split into a Dirichlet part \(\partial\Omega_D\) and Neumann part \(\partial\Omega_N\) with prescribed values \(g\) and derivatives \(h\) respectively. Here \(\Delta\) is the negative-semidefinite Laplacian, \(\sigma \in \mathbb{R}_{\geq 0}\) is a constant, and \(f\) is a given source term.  At interior points \(x\), the solution to \eqref{MainPDE} is given by the \emph{boundary integral equation (BIE)} \citep{Costabel:1987:BEM,Hunter:2001:BEM}
\begin{equation}
    \label{eq:BoundaryIntegralEquation}
    u(x)\! = \!\underbrace{\int_{\partial \Omega}\!\! \tfrac{\partial G}{\partial n}(x, z) u(z)\!-\!G(x, z) \tfrac{\partial u}{\partial n}(z) \ud z}_{\eqqcolon u_{\partial \Omega}(x)} + \!\underbrace{\int_{\Omega}\!\!G(x, y)\ f(y) \ud y}_{\eqqcolon u_{\Omega}(x)},
\end{equation}
where $G$ is the free-space Green's function for \eqref{MainPDE}, and $n$ is the unit outward normal at the boundary.

To make use of the BIE, one must somehow determine the unknown boundary data: Dirichlet values \(u\) on the Neumann boundary \(\partial\Omega_N\), and Neumann values \(\nicefrac{\partial u}{\partial n}\) on the Dirichlet boundary \(\partial\Omega_D\).  Schemes such as the \emph{boundary element method (BEM)} use a finite-dimensional space of functions on the boundary (\eg{}, basis functions associated with mesh nodes), and solve a dense, globally-coupled linear system for the best approximation to the true solution.

We take a completely different approach, and instead use random walks to compute the unknown boundary values.  In particular, we use WoS(t) to obtain \(u\) along \(\partial\Omega_N\) and \(\nicefrac{\partial u}{\partial n}\) along \(\partial \Omega_D\) (\secref{PointwiseEstimators}).  This approach avoids global solves, boundary remeshing, and approximation of the function space; unlike BEM, it also handles the source term \(f\).  Moreover, as random walks can be expensive (especially in problems with predominantly Neumann boundaries), we cache these boundary values at a collection of random sample points along \(\partial\Omega\).  We can then use a Monte Carlo estimate of \eqref{BoundaryIntegralEquation} to cheaply evaluate the solution at any interior point \(x\), without taking any further random walks (\algref{SampleReuse}).  This scheme is easy to parallelize, and can be computed progressively (\eg{}, for interactive preview).  We can also focus computation on a region of interest by caching points only on the boundary of a small subdomain \(R \subset \Omega\) (\secref{BoundarySpecification})---unlike BEM which must always perform a global solve involving the entire boundary \(\partial \Omega\).

In practice, we obtain far smoother results across the domain with our method compared to directly using pointwise estimators like WoS or WoSt (Figures \ref{fig:teaser}, \ref{fig:SplatterRMSE} and \ref{fig:Gradient}). This behavior can be attributed to correlations in the solution estimates at interior evaluation points that use the same boundary and source samples. On the flip side, error is now more global akin to traditional PDE solvers such as FEM and BEM (\figref{GlobalError}). Unlike pointwise estimators, we also observe boundary artifacts (\figref{BiasCorrection}) as samples are no longer generated in proportion to the singular functions $G$ and $\nicefrac{\partial G}{\partial n}$. We show how to mitigate such artifacts in \secref{MitigatingSingularArtifacts}.

\section{Related Work}\label{sec:RelatedWork}

Though we exclusively employ a grid-free Monte Carlo approach to solve PDEs such as the Poisson equation, numerous deterministic and stochastic solvers exist to accomplish this task with varying tradeoffs. We refer the reader to \citet[Section 7]{Sawhney:2020:MCGP} and \citet[Section 7]{Sawhney:2022:VCWoS} for a comprehensive overview of the advantages of a Monte Carlo approach like ours over traditional grid-based methods such as finite differences (FD), finite elements (FEM) and ``meshless'' finite elements (MFEM). Here we instead focus on techniques that directly evaluate BIEs, as well as sample reuse strategies in rendering our method is inspired by.

\paragraph{Grid-free Monte Carlo methods}
% \label{sec:GridFreeMonteCarloMethods}
Originally developed by \citet{Muller:1956:WOS}, WoS was recently introduced to graphics by \citet{Sawhney:2020:MCGP} who made the link to techniques from Monte Carlo rendering. It has since been generalized to handle Neumann boundary conditions \citep{Sawhney:2023:WalkOnStars}, solve variable coefficient PDEs \citep{Sawhney:2022:VCWoS}, simulate fluids \citep{Rioux-Lavoie:2022:MCFluid} and solve inverse problems via a differentiable formulation \cite{Yilmazer:2022:DiffWOS}. However, apart from proof-of-concept demonstrations \citep[Figures 12 \& 13]{Sawhney:2020:MCGP}, a practical scheme for denoising results has yet to be developed. \citet{Qi:2022:BidirWOS}'s recent bidirectional WoS formulation provides substantial variance reduction in problems with concentrated source terms, but is currently limited to only Dirichlet conditions and is not output sensitive---in contrast, our method can handle mixed boundary conditions and provides explicit control over the location and number of cached samples assigned to each evaluation point for estimation. As in rendering \citep{Segovia:2006:bidirectional}, it is likely that \citet{Qi:2022:BidirWOS}'s bidirectional approach can be used within our caching strategy to also estimate the solution at boundary samples.
%\todo{Cite WoB paper?}

\begin{figure}[t]
    \centering
    \includegraphics[width=\columnwidth]{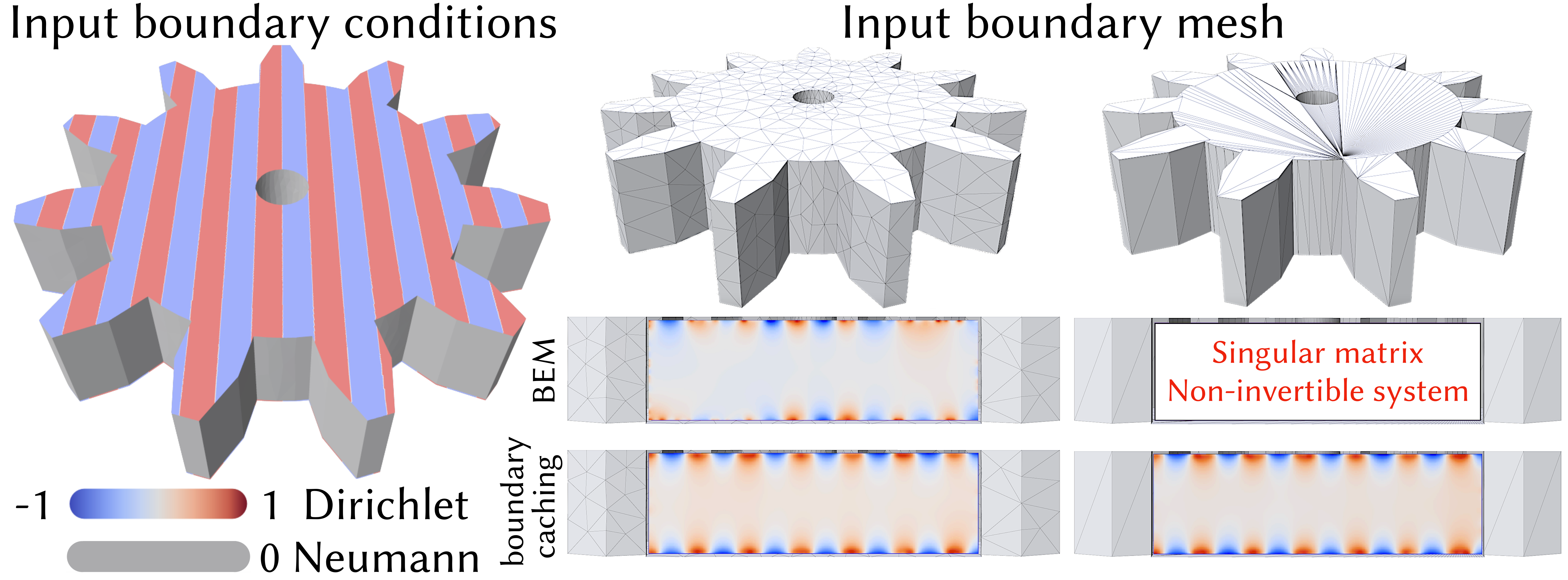}
    \caption{Even for problems with relatively simple boundary conditions and no source term (\emph{left}), finite element technology like BEM suffers from large global errors in the PDE solution without significant mesh refinement due to local aliasing of boundary data, and can fail completely on domains with irregular elements (\emph{middle row}). In contrast, our method solves PDEs without any aliasing artifacts irrespective of tesselation quality as it decouples problem inputs from the boundary representation (\emph{bottom row}).}
    \label{fig:BEM}
\end{figure}

\paragraph{Deterministic Boundary Element Solvers}\label{sec:DeterministicBoundaryElementSolvers}
Much like our grid-free PDE solver, the boundary element method (BEM) is designed to solve BIEs and does not discretize the interior of the domain. However, there are significant differences in capabilities: to evaluate \eqref{BoundaryIntegralEquation}, BEM must first discretize the boundary geometry, leading to spatial aliasing in the boundary data (\figref{BEM}). It must then invert a \emph{dense} linear system of equations for the unknown solution value $u$ on $\partial \Omega_N$ and normal derivative \(\nicefrac{\partial u}{\partial n}\) on $\partial \Omega_D$. As the linear system scales quadratically in size with the number of boundary elements, BEM does not truly reduce the dimensionality of a PDE solve compared to grid-based solvers that use sparse matrices, and requires special techniques like \emph{hierarchical matrix approximation}~\citep{Hackbusch:2015:Hierarchical} to achieve reasonable performance. BEM does not allow for progressive evaluation, nor is output-sensitive as $u$ and \(\nicefrac{\partial u}{\partial n}\) must always be determined on the entire boundary. Basic versions also ignore source terms---in this case, BEM must be coupled with a second interior solver such as FD, FEM and MFEM, inheriting shortcomings of grid-based solvers in the process \citep{Partridge:2012:DRBEM,Costabel:1987:BEM,Coleman:1991:EBE}. In contrast, our approach is progressive and output-sensitive (\secref{BoundarySpecification}). It is also scalable to increasing geometric detail, as the underlying pointwise estimators only require a spatial hierarchy to perform queries \citep[Section 5]{Sawhney:2023:WalkOnStars} and work with geometric representations beyond boundary meshes (\citep[Figure 2]{Sawhney:2020:MCGP} \& \citep[Figure 20]{Sawhney:2023:WalkOnStars}).

\begin{figure}[t]
    \centering
    \includegraphics[width=\columnwidth]{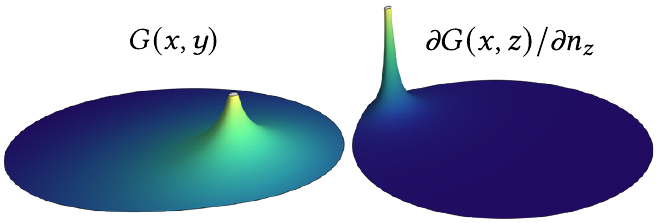}
    \caption{The free-space Green's function and its normal derivative in \eqref{BoundaryIntegralEquation} are singular at the point they are centered on, but decay smoothly and fall-off quickly away from the singularity.}
    \label{fig:FreeSpaceKernels}
\end{figure}

The (weakly) singular yet smoothly decaying Green's function and its normal derivative (\figref{FreeSpaceKernels}) in \eqref{BoundaryIntegralEquation} can be problematic to integrate near the boundary with both BEM and our method (\figref{BiasCorrection}). For BEM, specialized quadrature schemes can mitigate numerical issues due to singularities \citep{Fairweather:1998:MFS,Chen:2002:BKM,Chen:2009:SBM}. For us, not \emph{importance sampling} these functions leads to \emph{local} artifacts that adversely affect estimation variance. \secref{MitigatingSingularArtifacts} provides strategies to address this issue.

After computing $u$ and \(\nicefrac{\partial u}{\partial n}\) on the boundary with BEM, \eqref{BoundaryIntegralEquation} is often evaluated using fast multipole or Barnes-Hut schemes \citep{Greengard:1987:FMM,Pfalzner:1997:BH}. These acceleration strategies are often necessary due to the quadratic complexity of evaluating the BIE. We leave this optimization to future work as our current bottleneck is not BIE evaluation, but rather computing pointwise estimates on the boundary.

\paragraph{Sample Reuse in Rendering}\label{sec:SampleReuseRendering}
Popular sample reuse schemes in rendering such as virtual point lights \citep{Keller:1997:Instant, Dachsbacher:2014:Scalable}, photon mapping \citep{Jensen:1996:Global, Hachisuka:2009:Photon} and ReSTIR \citep{Bitterli:2020:ReSTIR, Ouyang:2021:ReSTIRGI} share samples across pixels to amortize the cost of long ray-traced paths and inject global information into per-pixel radiance estimates. As a result, they are often more efficient at rendering scenes with complex geometry and lighting than brute-force path tracers.

Our approach shares similarities with VPLs and photon mapping, in that samples generated and deposited on the scene (for us $\partial \Omega$) determine the flux (\ie, the solution $u$) over the image plane (which for us is either the entire domain $\Omega$ or a subset thereof). Unlike photon mapping however, our caching strategy does not require an additional data structure like a \emph{kd-tree} to store samples; instead we opt for a progressive formulation that discards boundary and source samples after splatting solution and gradient estimates in $\Omega$. Similarly to VPL methods, Monte Carlo noise is visually suppressed \citep[Figure 1]{Dachsbacher:2014:Scalable} from a combination of introducing correlations between estimates in $\Omega$, and the smooth decay of the functions $G$ and $\partial G/\partial n$ away from the boundary.
%($\partial G/\partial n$ in fact corresponds to the \emph{geometry term} in rendering; see \citet[Section 4.1]{Sawhney:2023:WalkOnStars}).
VPLs are also prone to singularities, as sharing samples requires sacrificing perfect importance sampling \citep[Section 5 \& Figure 9]{Dachsbacher:2014:Scalable}; our artifact correction schemes take inspiration from similar techniques for VPLs \citep{Kollig:2006:VPL}. Unlike VPL methods, we do not require testing for occlusion between deposited samples and evaluation points as the BIE contains no \emph{visibility term}.

Techniques based on \emph{lightcuts} \citep{Walter:2005:Lightcuts,Yuksel:2019:Lightcuts,Lin:2020:Lightcuts} can render scenes containing thousands of VPLs in real time. These methods cluster VPLs spatially in a tree, and then probabilistically select a subset of the VPLs that make the largest contribution at a given point. Akin to fast multipole and Barnes-Hut schemes, lightcuts would likely accelerate our method in domains with many evaluation points.

\begin{figure}[t]
    \centering
    \includegraphics[width=\columnwidth]{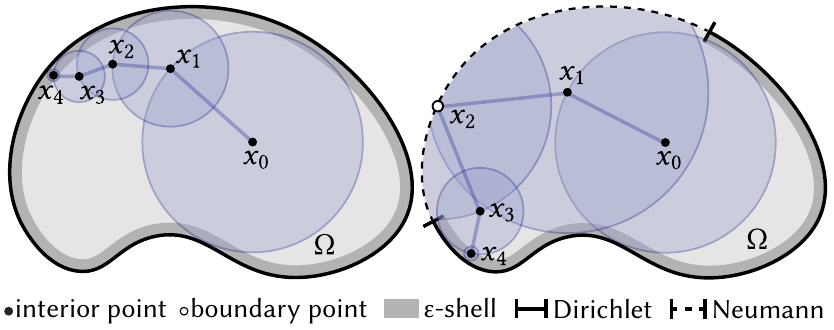}
    \caption{\emph{Left:} Walk on spheres simulates a \emph{Brownian} random walk inside a Dirichlet boundary $\partial \Omega_D$ to solve elliptic PDEs, repeatedly jumping to a random point on the largest sphere centered at the current walk location. The walk is terminated when it enters an epsilon shell $\partial \Omega_D^{\varepsilon}$ around the boundary. \emph{Right:} Walk on stars generalizes WoS to domains with a Neumann boundary $\partial \Omega_N$, using a sphere that can contain a subset of $\partial \Omega_N$ inside it. The next walk location is determined by intersecting a ray with a uniformly sampled direction against the sphere and parts of $\partial \Omega_N$ it contains, picking the first hit point. Like WoS, the walk terminates inside $\partial \Omega_D^{\varepsilon}$.}
    \label{fig:WoSvsWoSt}
\end{figure}

\section{Method}\label{sec:Method}
Our method estimates the solution $u = u_{\partial \Omega} + u_{\Omega}$ in \eqref{BoundaryIntegralEquation} at a set of evaluation points $\text{evalPts} \coloneqq \{x_k \in \Omega\}_{k=1}^K$ in a closed domain $\Omega \subset \R^N$ for, \eg, dense visualization of the solution. To this end, we create two caches, $\text{boundarySamples} \coloneqq \{z_n, \widehat{u}(z_n), \widehat{\nicefrac{\partial u}{\partial n}}(z_i)\}_{i=1}^N$ and $\text{sourceSamples} \coloneqq \{y_j, f(y_j)\}_{j=1}^M$, where $N$ and $M$ are user-specified cache sizes. The points $z_i$ and $y_j$ are sampled on the boundary $\partial \Omega$ and inside the domain $\Omega$ using probability densities $p^{\partial \Omega}$ and $p^{\Omega}$ respectively. The pointwise estimates $\widehat{u}(z_i)$ and $\widehat{\nicefrac{\partial u}{\partial n}(z_i)}$ are computed using WoS(t) (\secref{PointwiseEstimators}), while $f(y_j)$ are evaluations of the known source term.

Our method then uses the two caches to form \emph{correlated} Monte Carlo estimates
%\citep{fishman2006monte}
%\citep[Chapter 13]{Pharr:2016:PBR}
of \eqref{BoundaryIntegralEquation} at all evaluation points in $\text{evalPts}$:
\begin{align}
    \widehat{u_{\partial \Omega}}(x_k) &\coloneqq \frac{1}{N}\sum\nolimits_{i=1}^N \frac{\frac{\partial G}{\partial n}(x_k, z_i)\ \widehat{u}(z_i) - G(x_k, z_i) \widehat{\frac{\partial u}{\partial n}}(z_i)}{p^{\partial \Omega}(z_i)},\label{eq:BoundaryEstimator}\\
    \widehat{u_{\Omega}}(x_k) &\coloneqq \frac{1}{M}\sum\nolimits_{j=1}^M \frac{G(x_k, y_j)\ f(y_j)}{p^{\Omega}(y_j)}\label{eq:SourceEstimator}.
\end{align}
%
%where $y_i$ and $z_i$ are points sampled inside $\Omega$ and on its boundary $\partial \Omega$ using the probability densities $p^{\Omega}$ and $p^{\partial \Omega}$, respectively. Here, $N$ and $M$ are the number of samples we use to evaluate the boundary and source integrals. The source term $f$ is given as input, whereas we estimate $\widehat{u}(z_i)$ and $\widehat{\nicefrac{\partial u}{\partial n}}(z_i)$ using the WoSt and WoS algorithms we discuss next.
In \appref{GreensFns}, we provide expressions for $G$ and $\nicefrac{\partial G}{\partial n}$ for the Poisson and screened Poisson equations; \algref{SampleReuse} provides pseudocode, and \appref{BIEDoubleSided} discusses the extension to open domains and double-sided boundary conditions. In the rest of this section, we detail key aspects of our approach.
%We focus on the boundary sample cache $\text{boundarySamples}$, as the source sample cache is trivial to compute.

\begin{figure}[t]
    \centering
    \includegraphics[width=\columnwidth]{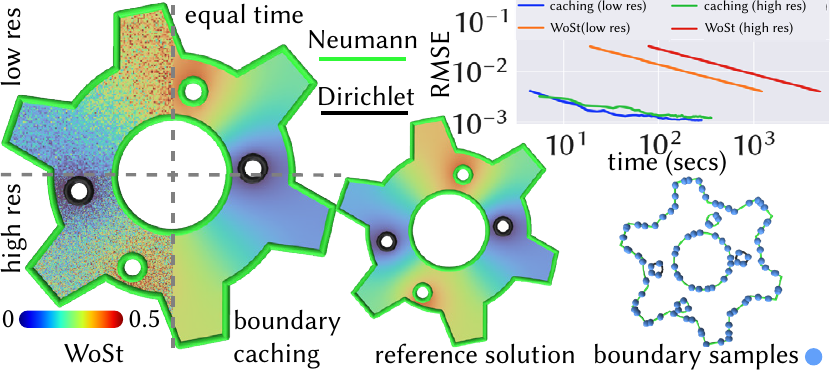}
    \caption{We obtain far smoother results across the domain compared to directly using pointwise estimators like WoSt at equal time. Evaluating the BIE at more points inside the domain has little impact on performance, as the same boundary samples are used to determine the PDE solution.}
    \label{fig:SplatterRMSE}
\end{figure}

\paragraph{Progressive Evaluation} Our method is progressive in two ways: First, we can improve estimation quality at a set of evaluation points, by generating new caches and using them to update existing estimates ($\Proc{UpdateSolution}$). Second, we can compute solution estimates at new evaluation points by iterating over existing caches.

\paragraph{Bias.} Assuming the pointwise estimates $\widehat{u}(z_i)$ and $\widehat{\nicefrac{\partial u}{\partial n}}(z_i)$ are unbiased, our estimator (Equations~\eqorigref{BoundaryEstimator} and~\eqorigref{SourceEstimator}) is also unbiased via the linearity of expectation. In reality, most pointwise estimators have a small amount of controllable bias, as is well established in the WoS literature (\secref{PointwiseEstimators}). Our caching scheme does not introduce any additional bias, while greatly improving efficiency (\figref{SplatterRMSE}).

\paragraph{Gradient Estimation}
Our method can reuse \emph{the same} cached boundary and source samples to also form Monte Carlo estimates of the solution \emph{gradient} at each evaluation point (\figref{Gradient}):
\begin{align}
    \widehat{\frac{\partial u_{\partial \Omega}}{\partial x}}\!(x_k)\! &\coloneqq\! \frac{1}{N}\!\sum\nolimits_{i=1}^N\!\!\frac{\tfrac{\partial^2 G}{\partial x \partial n}(x_k, z_i)\ \widehat{u}(z_i) - \tfrac{\partial G}{\partial x}(x_k, z_i) \widehat{\tfrac{\partial u}{\partial n}}(z_i)}{p^{\partial \Omega}(z_i)},\label{eq:BoundaryGradientEstimator}\\
    \widehat{\frac{\partial u_{\Omega}}{\partial x}}\!(x_k)\! &\coloneqq\! \frac{1}{M}\!\sum\nolimits_{j=1}^M\!\! \frac{\tfrac{\partial G}{\partial x}(x_k, y_j)\ f(y_j)}{p^{\Omega}(y_j)}.\label{eq:SourceGradientEstimator}
\end{align}
We provide expressions for the derivatives of $G$ in \appref{GreensFns}.

\begin{algorithm}[t]
\caption{A cache-based strategy to evaluate the BIE at points inside a closed user-defined boundary $\partial R$ (\secref{BoundarySpecification})}
\label{alg:SampleReuse}
\begin{algorithmic}[1]
\algblockdefx[Name]{Struct}{EndStruct}
    [1][Unknown]{\textbf{struct} #1}
    {}
\algtext*{EndStruct}
\algblockdefx[Name]{FORDO}{ENDFORDO}
    [1][Unknown]{\textbf{for} #1 \textbf{do}}
    {}
\algtext*{ENDFORDO}
\algblockdefx[Name]{PARALLELFORDO}{ENDPARALLELFORDO}
    [1][Unknown]{\textbf{parallel}\ \textbf{for} #1 \textbf{do}}
    {}
\algtext*{ENDPARALLELFORDO}
\algblockdefx[Name]{IF}{ENDIF}
    [1][Unknown]{\textbf{if} #1 \textbf{then}}
    {}
\algtext*{ENDIF}
\algblockdefx[Name]{IFTHEN}{ENDIFTHEN}
    [2][Unknown]{\textbf{if} #1 \textbf{then} #2}
    {}
\algtext*{ENDIFTHEN}
\algblockdefx[Name]{RETURN}{ENDRETURN}
    [1][Unknown]{\textbf{return} #1}
    {}
\algtext*{ENDRETURN}
\algblockdefx[Name]{COMMENT}{ENDCOMMENT}
    [1][Unknown]{\textcolor{commentpaleblue}{\(\triangleright\)\textit{#1}}}
    {}
\algtext*{ENDCOMMENT}

\Struct[$\Proc{BoundarySample}$]
    \State $z, n_z \gets \textsc{null}$\Comment{Sample location \& unit outward normal on $\partial R$}
    \State $\widehat{u}, \widehat{\frac{\partial u}{\partial n}} \gets 0$\Comment{Estimates for solution \& normal derivative}
\EndStruct
\Struct[$\Proc{EvaluationPoint}$]
    \State $x \gets \textsc{null}$\Comment{Location for evaluating BIE (\eqref{BoundaryIntegralEquation})}
    \State $\widehat{u}_{\partial \Omega}^{\texttt{sum}}, \widehat{u}_{\Omega}^{\texttt{sum}} \gets 0$\Comment{Running sums for solution evaluation}
    \State $N, M \gets 0$\Comment{boundary \& source sample count}
    \Function{GetSolution}{}
        \RETURN[$\widehat{u}_{\partial \Omega}^{\texttt{sum}} \big/ N\ +\ \widehat{u}_{\Omega}^{\texttt{sum}} \big/ M$]
        \ENDRETURN
    \EndFunction
\EndStruct
\Require A set of evaluation points $\texttt{evalPts}$, cache sizes $N$ \& $M$ for boundary and source samples, number of walks $\texttt{nWalks}$ to take for pointwise estimation, and densities $p^{\partial R}$ \& $p^{R}$ for sample generation.
\Ensure An updated solution estimate for each evaluation point.
\Function{UpdateSolution}{$\texttt{evalPts},\ N,\ M,\ \texttt{nWalks},\ p^{\partial R},\ p^{R}$}
    \COMMENT[Generate $N$ boundary samples and estimate $u$ \& \(\nicefrac{\partial u}{\partial n}\) for each]\ENDCOMMENT
    \State $\texttt{boundarySamples}\ \gets \Proc{BoundarySample}[N]$
    \PARALLELFORDO[$b\ \textbf{in}\ \texttt{boundarySamples}$]
        \State $b.z, b.n_z \sim p^{\partial R}$
        \State $b.\widehat{u}, b.\widehat{\frac{\partial u}{\partial n}} \gets \Proc{WalkOnStars}(b.z, b.n_z, \texttt{nWalks})$\Comment{Sec. \ref{sec:PointwiseEstimators}}
    \ENDPARALLELFORDO

    \COMMENT[Splat contributions from boundary samples to $\texttt{evalPts}$]\ENDCOMMENT
    \FORDO[$b\ \textbf{in}\ \texttt{boundarySamples}$]
        \State $z, n_z \gets b.z, b.n_z$
        \PARALLELFORDO[$e\ \textbf{in}\ \texttt{evalPts}$]
            \State $e.\widehat{u}_{\partial \Omega}^{\texttt{sum}} \mathrel{+}= \left(\frac{\partial G}{\partial n}(e.x,\ z) b.\widehat{u} - G(e.x, z) b.\widehat{\frac{\partial u}{\partial n}}\right)\big/ p^{\partial R}(z)$
            \State $e.N \mathrel{+}= 1$
        \ENDPARALLELFORDO
    \ENDFORDO

    \COMMENT[Generate $M$ source samples \& splat their contributions to $\texttt{evalPts}$]\ENDCOMMENT
    \FORDO[$j\ \textbf{in}\ \Proc{Range}(M)$]
        \State $y \sim p^{R}$
        \PARALLELFORDO[$e\ \textbf{in}\ \texttt{evalPts}$]
            \State $e.\widehat{u}_{\Omega}^{\texttt{sum}} \mathrel{+}= G(e.x,\ y)\ f(y) \big/ p^{R}(y)$
            \State $e.M \mathrel{+}= 1$
        \ENDPARALLELFORDO
    \ENDFORDO
\EndFunction
\end{algorithmic}
\end{algorithm}

\newcommand{\epsilonShellFigure}{%
\setlength{\columnsep}{1em}
\setlength{\intextsep}{0em}
\begin{wrapfigure}[13]{r}{88pt}
    \centering
    \includegraphics{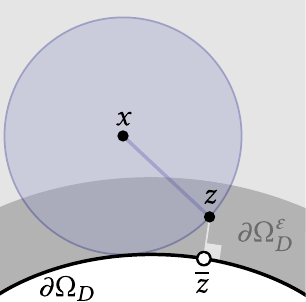}
    \caption{Inside $\partial \Omega_D^{\varepsilon}$, WoS(t) uses the known Dirichlet data $g$ from the closest projected point on $\partial \Omega_D$.}
    \label{fig:EpsilonShell}
\end{wrapfigure}%
}
\subsection{The WoS \& WoSt Estimators}\label{sec:PointwiseEstimators}
%Our caching strategy uses WoSt to compute the pointwise solution estimate $\widehat{u}$ in \eqref{BoundaryEstimator} for any boundary sample $z_i$. To estimate $\partial u/\partial n$, we employ the gradient estimation strategy for WoS proposed in \citet[Section 3]{Sawhney:2020:MCGP}, which for any point $x \in \Omega$ uses a ball $B(x)$ contained entirely in $\Omega$ to estimate the solution gradient at $x$:
We use the WoS and WoSt algorithms to compute pointwise estimates at each boundary sample $z_i$. WoS \citep{Muller:1956:WOS} is a Monte Carlo algorithm that recursively evaluates the \emph{mean-value property} of harmonic functions to solve elliptic PDEs in a bounded domain $\Omega$ with Dirichlet boundary conditions. It does so by repeatedly jumping\epsilonShellFigure{}to a random point on the \emph{largest} sphere centered at the current random walk location (\figref{WoSvsWoSt}, \emph{left}). WoSt \citep{Sawhney:2023:WalkOnStars} generalizes WoS to solve problems with mixed Dirichlet and Neumann boundary conditions: instead of spheres wholly-contained in $\Omega$, WoSt uses \emph{star-shaped regions} formed by intersecting spheres with parts of the Neumann boundary $\partial \Omega_N$
%, to simulate reflecting random walk
(\figref{WoSvsWoSt}, \emph{right}). Both algorithms terminate walks when they enter the \emph{epsilon shell} $\partial \Omega_D^{\varepsilon}$ around the Dirichlet boundary (inset), where they use the known boundary value $g$ at the nearest boundary point. This introduces a small bias into the solution estimate $\widehat{u}$, which diminishes at the rate $O(\nicefrac{1}{\log \varepsilon})$ \citep{Binder:2012:WosRate};
%shrinking $\varepsilon$ has little impact on performance \citep[Section 6.1]{Sawhney:2020:MCGP}.
we set $\varepsilon$ to be $0.001 \times$ the diagonal length of the scene.

Both WoS and WoSt replace the free-space Green's function $G$ in \eqref{BoundaryIntegralEquation} with the Green's function $G^B$ for a ball $B(x)$ centered at $x$---this choice simplifies the BIE and facilitates recursive estimation as discussed in \citet[Sections 3 \& 4]{Sawhney:2023:WalkOnStars}.
%---in particular, these pointwise estimators perform recursive random walks to estimate just the unknown solution value $u$ in the BIE as the dependence on the unknown $\partial u/\partial n$ values drops with $G^B = 0$ on $\partial B(x)$ (see \citet[Sections 3 \& 4]{Sawhney:2023:WalkOnStars}).
Importantly for our method, the use of $G^B$ makes it possible to also estimate the normal derivative $\nicefrac{\partial u}{\partial n}$ in \eqref{BoundaryIntegralEquation}. This is done through recursive estimation of the following integral expression at any $x \in \Omega$ \citep[Section 3]{Sawhney:2020:MCGP}:
\begin{equation}
    \label{eq:SolutionGradientBall}
    % \frac{\partial u}{\partial x}(x)\! =\! \underbrace{\int_{\partial B(x)}\!\! \frac{\partial^2 G^B}{\partial x \partial n}\!(x, z) u(z) \ud z}_{\eqqcolon \nicefrac{\partial u_{\partial \Omega}}{\partial x}(x)} +\! \underbrace{\int_{B(x)}\!\! \frac{\partial G^{B}}{\partial x}(x, y) f(y) \ud y}_{\eqqcolon \nicefrac{\partial u_{\Omega}}{\partial x}(x)}.
    \frac{\partial u}{\partial x}(x)\! =\! \int_{\partial B(x)}\!\! \frac{\partial^2 G^B}{\partial x\ \partial n}\!(x, z)\ u(z) \ud z +\! \int_{B(x)}\!\! \frac{\partial G^{B}}{\partial x}(x, y)\ f(y) \ud y,
\end{equation}
where the ball $B(x)$ is selected as in WoS, and the unknown value $u(z)$ inside $\partial B(x)$ is recursively estimated using WoS for pure Dirichlet problems, and WoSt for mixed boundary-value problems. Given a unit normal $n_x$ at $x$, the normal derivative is then estimated using $n_x \cdot \widehat{\nicefrac{\partial u}{\partial x}}(x)$. As with $\widehat{u}$, this estimate is slightly biased due to the epsilon shell used for termination. We refer to \citet[Equation 13 \& Appendix B.1]{Sawhney:2020:MCGP} for derivatives of $G^B$, and to \citet[Section 4.1.1]{Sawhney:2020:MCGP} and \citet[Section 4.1]{Rioux-Lavoie:2022:MCFluid} for variance reduction strategies.
%when $B(x)$ intersects the epsilon shell $\partial \Omega_D^{\varepsilon}$ as the unknown solution $u(z)$ in \eqref{SolutionGradientBall} for any point $z \in \partial \Omega_D^{\varepsilon}$ is set to the known Dirichlet data at the closest projected point $\overline{z} \in \partial \Omega_D$ (\figref{EpsilonShell}).

\begin{figure}[t]
    \centering
    \includegraphics[width=\columnwidth]{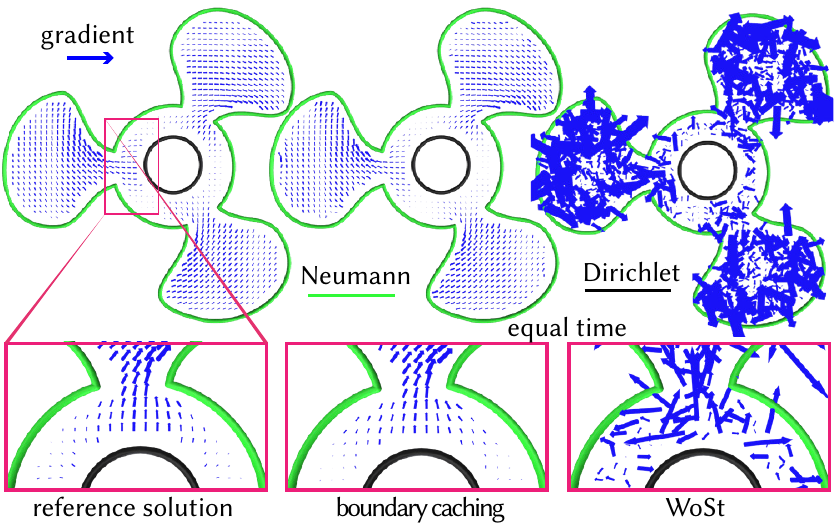}
    \caption{Our gradients have considerably less noise compared to pointwise estimates as they use known values for \(\nicefrac{\partial u}{\partial n}\) on the Neumann boundary to evaluate \eqref{BoundaryGradientEstimator}. In contrast, WoSt gradients become noisier away from the Dirichlet boundary as estimation requires longer random walks.}
    \label{fig:Gradient}
\end{figure}

\subsection{Boundary Specification}\label{sec:BoundarySpecification}
\begin{figure}[t]
    \centering
    \includegraphics[width=\columnwidth]{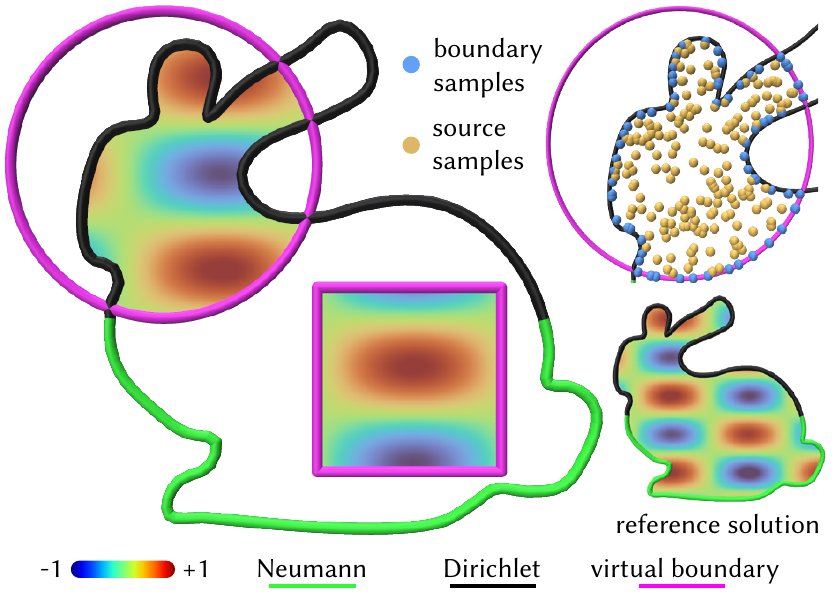}
    \caption{Unlike BEM which must always perform a global solve involving the entire boundary, we can focus computation on a region of interest by caching points only on the boundary of smaller subdomains.}
    \label{fig:VirtualBoundary}
\end{figure}

When the solution needs to be evaluated within a localized region $R$ inside the domain $\Omega$ (\figref{VirtualBoundary}), we specialize the BIE to this region by generating source samples in $R$ and boundary samples on $\partial R$ (\algref{SampleReuse}, \emph{lines} 15 \& 25). We use uniform densities $p^{R} = \nicefrac{1}{|R|}$ and $p^{\partial R} = \nicefrac{1}{|\partial R|}$ for sample generation, though we could use densities specialized to specific PDEs and their inputs to reduce variance.
%The solution $u$ and its normal derivative $\nicefrac{\partial u}{\partial n}$ for each boundary sample are estimated using the pointwise estimators from the previous section; the unit outward normal to $\partial R$ is used to compute $\widehat{\nicefrac{\partial u}{\partial n}}$ (\algref{SampleReuse}, \emph{lines} 15 \& 16).
The solution integrates to 0 by construction outside $R$.

\begin{figure}[t]
    \centering
    \includegraphics[width=\columnwidth]{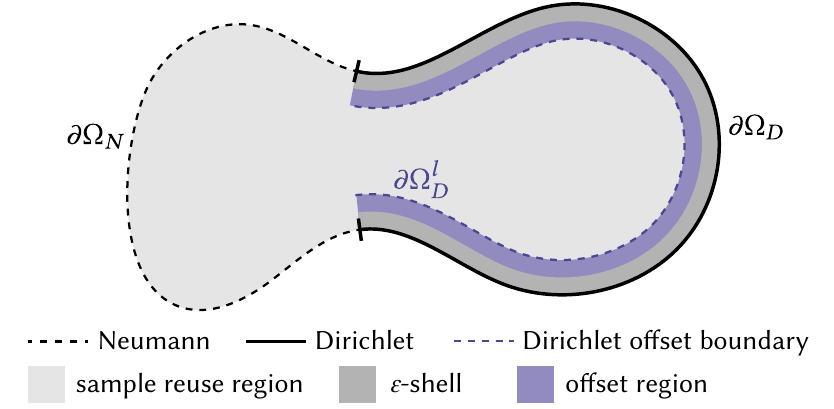}
    \caption{To solve the BIE inside $\Omega$, we generate samples on the Neumann boundary $\partial \Omega_N$ and an offset Dirichlet boundary $\partial \Omega_D^l$ where $l > \varepsilon$.}
    \label{fig:DirichletOffsetBoundary}
\end{figure}

When the solution needs to be evaluated within the entire domain $\Omega$ (\ie, $R \coloneqq \Omega$), we incorporate the known boundary data $\nicefrac{\partial u}{\partial n} = h$ on $\partial \Omega_N$ directly into our sample estimates, rather than estimating it from scratch. Unfortunately, estimating $\nicefrac{\partial u}{\partial n}$ on the Dirichlet boundary $\partial \Omega_D$ is challenging, as WoS uses a ball with a non-zero radius to estimate the solution gradient (\eqref{SolutionGradientBall}). Instead of dealing with $\partial \Omega_D$ directly, we define a closed region bounded by the Neumann boundary $\partial \Omega_N$ and an offset Dirichlet boundary $\partial \Omega_D^l$ where $l > \varepsilon$ (\figref{DirichletOffsetBoundary}). We then generate boundary samples on $\partial \Omega_D^l$, and estimate both $u$ and $\nicefrac{\partial u}{\partial n}$. Moreover, for any evaluation point that is within a distance $l$ to $\partial \Omega_D$, we use WoS(t) to compute the solution there, as random walks typically terminate quickly when started close to the Dirichlet boundary.

There are two considerations involved in choosing an offset $l$: the minimal feature size of the domain $\Omega$, and the amount of bias in the estimate for $\nicefrac{\partial u}{\partial n}$ on $\partial \Omega_D^l$ based on its proximity to the epsilon shell $\partial \Omega_D^{\varepsilon}$. We do not need to define sample reuse regions in the vicinity of thin features---instead, we can create multiple disconnected regions inside $\Omega$ where boundary and source samples are cached, while using WoSt to estimate the solution pointwise elsewhere. We set $l$ to $5 \times \varepsilon$ in our results (\figref{Ablations}, \emph{top} shows an ablation) to mitigate bias in the estimates for $\nicefrac{\partial u}{\partial n}$ on $\partial \Omega_D^l$, and leave principled and unbiased estimation of this quantity directly on $\partial \Omega_D$ to future work.

\begin{figure}[t]
    \centering
    \includegraphics[width=\columnwidth]{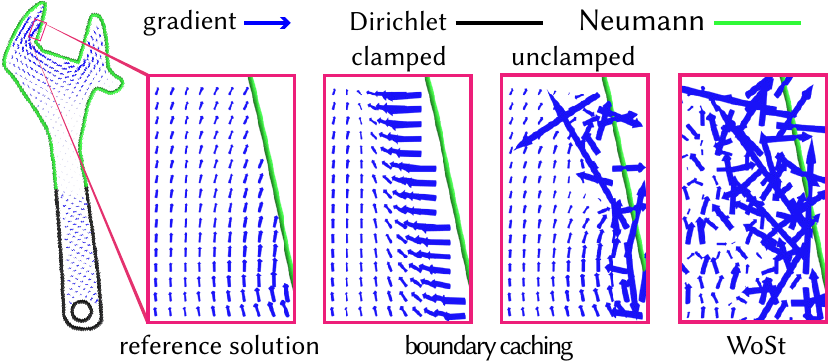}
    \caption{Na\"ively clamping singular functions in our gradient estimates suppresses noise near the boundary, but produces biased results.}
    \label{fig:GradientBoundary}
\end{figure}

\subsection{Singularities}\label{sec:MitigatingSingularArtifacts}
\begin{figure}[t]
    \centering
    \includegraphics[width=\columnwidth]{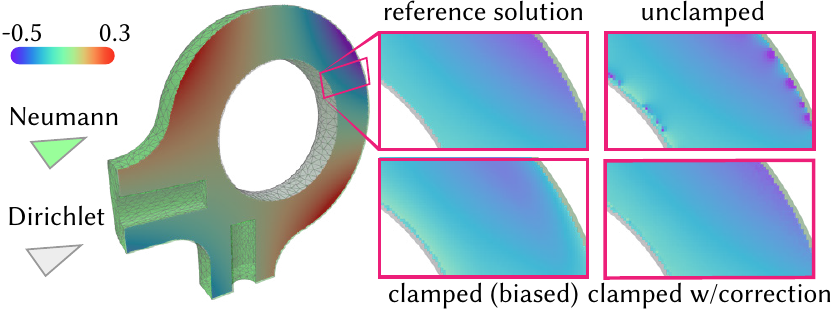}
    \caption{Our clamping strategy (\emph{bottom right}) effectively suppresses local artifacts from singular functions near the boundary without bias (\emph{top right}).}
    \label{fig:BiasCorrection}
\end{figure}

Though the free-space Green's function and its derivatives decay smoothly, they are singular at the point they are centered on (\figref{FreeSpaceKernels}). Therefore, just as in BEM, our basic method can suffer from local artifacts---especially near the boundary (\figref{BiasCorrection})---as it uses uniformly distributed source and boundary samples to evaluate \eqsref{BoundaryEstimator}{SourceGradientEstimator}. In contrast, such artifacts do not arise in WoS(t) as the corresponding functions for a ball are importance sampled, \ie, $p^{B} \propto G^B$ and $p^{\partial B} \ \propto \nicefrac{\partial G^B}{\partial n}$ \citep[Section 4]{Sawhney:2023:WalkOnStars}.

Similar artifacts are often suppressed in virtual point light methods by clamping the singular geometry term in the rendering equation \citep[Section 5 \& Figure 9]{Dachsbacher:2014:Scalable}. However, clamping can introduce noticeable bias in the solution estimate with both VPLs and our method (\figref{BiasCorrection}). Likewise, clamping derivative norms in \eqsref{BoundaryGradientEstimator}{SourceGradientEstimator} leads to smoothly-varying but biased gradients near the boundary (\figref{GradientBoundary}). Below, we extend \citet{Kollig:2006:VPL}'s bias correction strategy for VPLs to the function $\nicefrac{\partial G}{\partial n}$; in \appref{GreensFnArtifacts} we do the same for $G$, and leave the extension to gradient estimators to future work.

\paragraph{Removing Bias from Clamping} The function $\nicefrac{\partial G}{\partial n}(x, z)$ is large when the points $x$ and $z$ are close. To mitigate artifacts that arise from not importance sampling $\nicefrac{\partial G}{\partial n}$ for $z$, we rewrite the first term in \eqref{BoundaryIntegralEquation} over a user-defined region $R$ as follows:
\begin{equation}
    \label{eq:PoissonKernelIntegralClamped}
    \int_{\partial R} \left.\frac{\partial G}{\partial n}\right\rvert_c(x, z) u(z) \ud z\ +\ \int_{\partial R} \left[\frac{\partial G}{\partial n}(x, z) - \left.\frac{\partial G}{\partial n}\right\rvert_c(x, z)\right] u(z) \ud z,
\end{equation}
where $c$ is a positive user-specified bound, and
\begin{equation}
    \label{eq:PoissonKernelClamped}
    \left.\nicefrac{\partial G}{\partial n}\right\rvert_c \equiv \max(-c,\ \min(c,\ \nicefrac{\partial G}{\partial n})).
\end{equation}
As before, we use uniformly distributed boundary samples on $\partial R$ to estimate the first integral in \eqref{PoissonKernelIntegralClamped} at any evaluation point $x \in R$. The bound $c$ is set based on the scale of the scene (\figref{Ablations}, \emph{bottom} shows an ablation), though strategies for automatically setting this parameter can likely be adapted from the VPL literature \citep[Section 2.3]{Kollig:2006:VPL}.

To estimate the second integral, we importance-sample new boundary samples on $\partial R$ using the probability density $p^{\partial R} = \nicefrac{\partial G}{\partial n}$. For Poisson-like equations, this density defines the \emph{signed solid angle} over the boundary (\eqref{FreeSpacePoissonKernels}), and can be sampled exactly by intersecting a ray (with a uniformly sampled direction on the unit sphere) with $\partial R$ \citep[Section 5.5.3]{Pharr:2016:PBR}. We use WoSt to estimate the unknown solution value $u(z)$ at intersection points (which might be more than one if $\partial R$ is nonconvex; see \citet[Section 4.3]{Sawhney:2023:WalkOnStars}). However, we only need to estimate $u$ when $\left.\nicefrac{\partial G}{\partial n}\right\rvert_c(x, z) \neq \nicefrac{\partial G}{\partial n}(x, z)$, \ie, when the evaluation point $x$ is in close proximity to an intersection point $z$. This is typically not the case for most evaluation points inside $R$, as $\nicefrac{\partial G}{\partial n}$ falls off quickly (\figref{FreeSpaceKernels}). As a result, we can estimate \eqref{PoissonKernelIntegralClamped} in both an artifact- and bias-free manner by running a small number of random walks for the second integral; \figref{BiasCorrection} shows improvements provided by this approach.

\section{Implementation and Evaluation}\label{sec:ImplementationEvaluation}
\paragraph{Setup} We encode a problem instance by a description of the domain boundary $\partial \Omega$ and the functions $f$, $g$ and $h$ in \eqref{MainPDE}. Unlike grid-based solvers such as FEM and BEM, these functions are not discretized or approximated in a finite basis, and are instead provided via arbitrary callback routines that return a value for any query point $x \in \Omega$. We use boundaries represented as polygonal meshes in our experiments, and a CPU-based axis aligned \emph{bounding volume hierarchy (BVH)} to perform ray intersection and closest point queries needed by WoS(t) \citep[Section 5]{Sawhney:2023:WalkOnStars}. Relative to finite element mesh generation, a BVH uses little memory and is typically a lot faster to build even for detailed models \citep{Sawhney:2020:MCGP,Sawhney:2022:VCWoS,Sawhney:2023:WalkOnStars}.

\paragraph{Sampling} We use a discrete \emph{cumulative density function (CDF)} table \citep[Section 13.3]{Pharr:2016:PBR} with stratified random numbers to generate boundary samples over elements (\eg, triangles) of a polygonal mesh. Samples are drawn uniformly on elements selected in proportion to their surface area. Faster sample generation is possible with an alias table \citep{Walker:1974:AliasTable, Walker:1977:AliasTable}. Source samples are likewise generated uniformly inside $\Omega$ or a user-defined region $R$ with stratified sampling \citep[Section 13.8]{Pharr:2016:PBR}. We use this sampling setup for all figures in the paper except \figref{teaser}, where we find that results improve significantly if the boundary samples in Equations~\eqorigref{BoundaryEstimator} and~\eqorigref{BoundaryGradientEstimator} are weighted by the area of their associated Voronoi cell. Similar area weighting strategies have proven effective in surface reconstruction applications \citep[Figure 5]{Barill:2018:FWN}, and yield a consistent Monte Carlo estimator that provides provably better convergence \citep{Guo:2021:Geometric}.

\begin{figure}[t]
    \centering
    \includegraphics[width=\columnwidth]{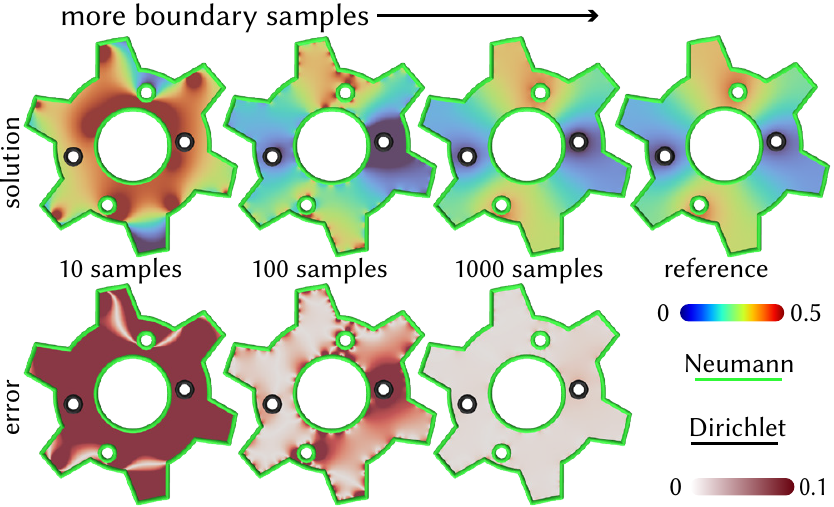}
    \caption{Akin to traditional PDE solvers, our method demonstrates a global error in the solution profile that vanishes with more boundary samples.}
    \label{fig:GlobalError}
\end{figure}

\paragraph{Convergence} In Figures \ref{fig:SplatterRMSE} and \ref{fig:Gallery}, we show that even in 2D our method provides up to an order of magnitude error reduction at equal time compared to WoSt for mixed boundary value problems. The reason is twofold: first, run-time efficiency improves as we do not perform independent random walks for interior evaluation points. Second, boundary samples inject global information about the solution into interior estimates. The overall result is that our solver reduces the dimensionality of a PDE solve---in contrast, BEM trades off discretizing the domain with having to solve a much denser linear system.

Similarly to traditional PDE solvers requiring global solves, error in the interior now depends on the number of boundary and source samples used. Figures \ref{fig:GlobalError} and \ref{fig:Gallery} show that error vanishes with more samples, though we note that \eqsref{BoundaryEstimator}{SourceGradientEstimator} provide unbiased estimates even with just a single sample. We also observe lower interior error with more accurate estimates for $u$ and $\nicefrac{\partial u}{\partial n}$ on the boundary. As derivative estimates are typically noisier than estimates of the solution \citep[Figure 15]{Sawhney:2020:MCGP}, we perform more random walks for the estimate $\widehat{\nicefrac{\partial u}{\partial n}}$ on an offset Dirichlet boundary $\partial \Omega_D^l$ (\secref{BoundarySpecification}) compared to $\widehat{u}$ on $\partial \Omega_N$. As a default, we take $10\times$ more walks for boundary samples on $\partial \Omega_D^l$ than on $\partial \Omega_N$. In general, the overhead is small as walks starting close to the Dirichlet boundary are usually much shorter.

For pure Dirichlet problems, we found empirically that the bidirectional WoS approach by \citet{Qi:2022:BidirWOS} can be more effective at equal time than our more general caching strategy (\figref{BidirectionalComparison})---by specializing to Dirichlet conditions, this approach avoids the need to estimate $\nicefrac{\partial u}{\partial n}$ on the boundary or evaluate $\nicefrac{\partial G}{\partial n}$ during BIE estimation, which leads to lower variance.

\begin{figure}[t]
    \centering
    \includegraphics[width=\columnwidth]{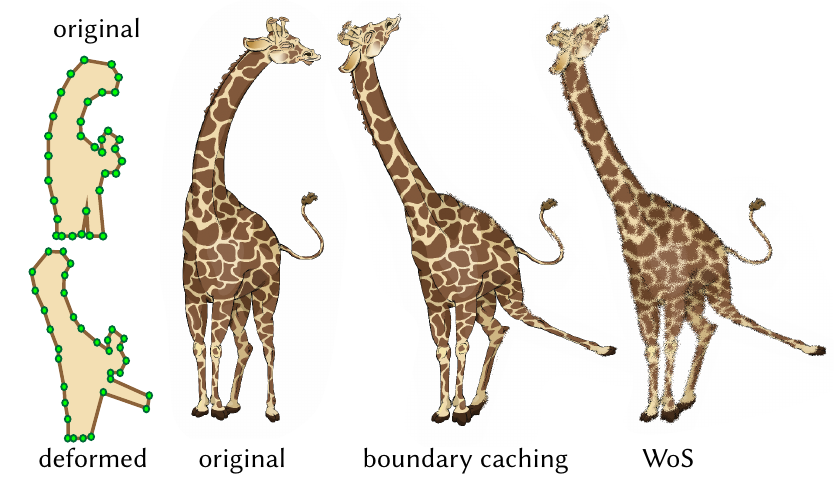}
    \caption{Compared to WoS, our caching strategy provides a noticeably smoother harmonic interpolation of texture coordinates in a deformed cage.}
    \label{fig:HarmonicDeformation}
\end{figure}

\begin{figure}[t]
    \centering
    \includegraphics[width=\columnwidth]{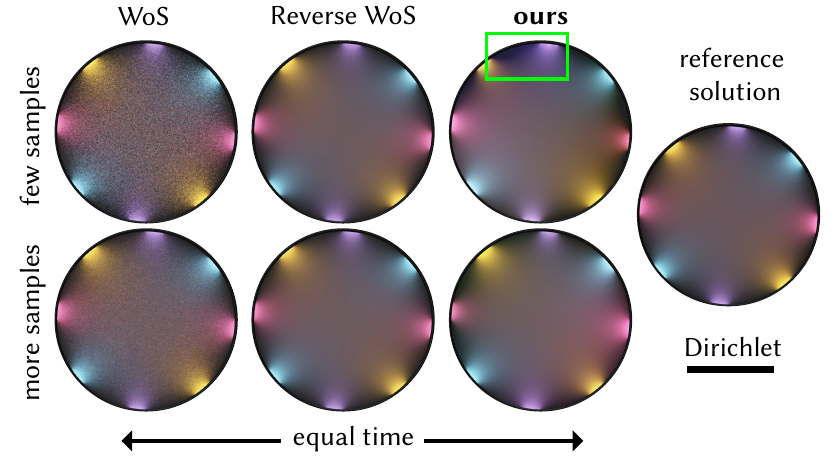}
    \caption{For Dirichlet problems, our method often has higher variance than \citet{Qi:2022:BidirWOS}'s specialized approach for these boundary conditions.}
    \label{fig:BidirectionalComparison}
\end{figure}

\paragraph{Test Problems} Given the spatial smoothness of solutions to elliptic problems, a key strength of our method is its ability to suppress the salt-and-pepper noise characteristic of independent Monte Carlo estimates. In \figref{HarmonicDeformation}, we solve a Laplace equation with pure Dirichlet boundary conditions to interpolate texture coordinates in a deformed cage. Our correlated sample estimates provide noticeably smoother results, suitable for visualization. We observe similar behavior with our gradient estimates in \figref{teaser}, where we solve for streamlines of a potential flow with Dirichlet boundary conditions of -1 and 1 at the front and back of the tunnel respectively, and 0 Neumann conditions elsewhere. Similarly to \citet[Section 5.2.6]{Sawhney:2020:MCGP}, we draw streamlines along the estimated gradient direction by numerically integrating an ordinary differential equation using Huen's method. Steamlines start at a collection of random seed points in the domain, and can be regenerated cheaply once estimates for $u$ and $\nicefrac{\partial u}{\partial n}$ have been computed on the boundary---we do not require a high-quality simulation mesh for this task.

\begin{figure*}
    \includegraphics{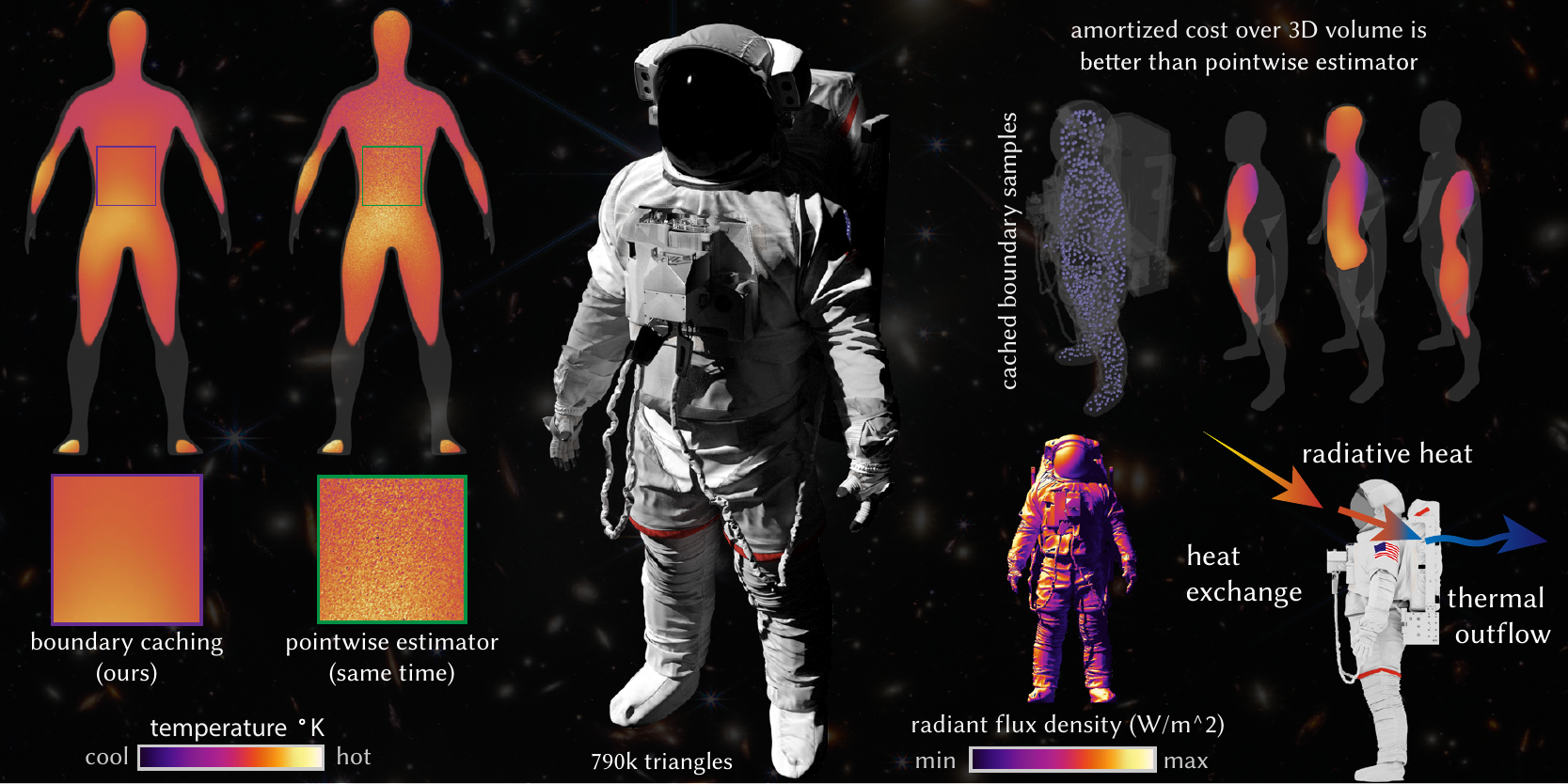}
    \caption{Regulating the body temperature of astronauts during extra-vehicular activities is a crucial function of spacesuits. Inspired by this application, we build a simplified physical model involving heat exchange (via Dirichlet conditions) and outflow (via Neumann conditions) on a spacesuit with an interior source term to model an astronaut's ambient body temperature.
    \label{fig:Spacesuit}
    }
\end{figure*}

In \figref{Spacesuit}, we demonstrate the ability of our solver to resolve detailed boundary conditions and source terms without the need for any volumetric mesh generation. In particular, we model how effectively a spacesuit regulates an astronaut's body temperature. We use a \emph{ray-traced} solar radiation simulation for Dirichlet conditions; Neumann conditions model outflow; the source term provides the ambient body temperature. Compared to pointwise estimators, our solver provides low-variance results on a high-resolution slice plane at equal time. Moreover, efficiency improves drastically when evaluating inside a dense 3D grid since we can reuse cached samples. In this case however, evaluating the BIE at all grid points is more expensive than estimating $u$ and $\nicefrac{\partial u}{\partial n}$ on the boundary.

%Despite their numerous strengths, pointwise estimators scale poorly with respect to the number of evaluation points. This scaling becomes a serious limitation for both dense solution grids and solutions in higher dimensions. Unlike rendering, it's common to solve a PDE everywhere in a domain. In \figref{Spacesuit}, we demonstrate how boundary caching can be leveraged to handle a dense set of evaluation points without sacrificing the strengths of a grid-free Monte Carlo estimator. We solve a thermal modeling problem for a localized region of interest and define a detailed Dirichlet boundary condition generated via a ray traced, solar radiation simulation. The PDE additionally contains a constant negative Neumann boundary condition on the back of the spacesuit to capture heat outflow and a spatially varying source analogous to body heat dissipation. As a result of sharing walks, boundary caching can in many cases consider orders of magnitudes more walks per each point. The cached boundary samples can additionally be used for evaluation of new sets of points at the cost of having to repeat the step of accumulating contributions. While this "splatting" step is a significantly less expensive operation than taking a full random walk, boundary caching's ability to amortize the cost of walks eventually forces splatting to become the primary bottleneck. We have yet to incorporate Barnes-Hut or other hierarchical evaluation methods to address this bottleneck, but it is a natural future direction.

\section{Conclusion and Future Work}\label{sec:ConclusionAndFutureWork}
We presented a simple caching strategy for an emerging class of PDE solvers based on grid-free Monte Carlo methods. Our method greatly improves the efficiency of densely evaluating a PDE solution and its gradient inside a domain (or a subset thereof), by performing random walks only on the boundary. It also significantly suppresses Monte Carlo noise without adversely affecting scalability, progressivity, and output sensitivity of the underlying pointwise estimators.

Looking ahead, the ability of our solver to locally evaluate solutions inside arbitrary bounding regions opens up the possibility of developing domain decomposition strategies \citep{Chan:1994:DD} in the Monte Carlo framework that more effectively handle domains with, \eg, thin features where pointwise estimators struggle due to long random walk lengths (see \citet[Section 7]{Sawhney:2023:WalkOnStars}). Likewise, our method offers opportunities for improving performance of the pointwise estimators it uses by terminating walks in regions where boundary values have previously been cached.

Our method can be improved in a number of ways. In particular, faster and less noisy WoS(t) estimators will not just improve the run-time performance of our approach, they will also reduce the global error in the estimated solution inside the domain. Error will likely reduce noticeably with a principled and unbiased technique for pointwise estimation of the normal derivative of the solution on the Dirichlet boundary. Variance can be improved by adaptively generating source and boundary samples from probability densities specialized to specific PDEs and their inputs. Moreover, though sharing samples necessitates a lack of importance sampling, taking further inspiration from VPL methods \citep[Section 5]{Dachsbacher:2014:Scalable} for removing singular artifacts will improve the quality of our results near the boundary, especially for gradient estimates. The quadratic complexity of evaluating the BIE can also be greatly accelerated in situations where estimation is needed at numerous points in the domain, by adapting and incorporating clustering techniques such as Barnes-Hut and lightcuts into our method.

Finally, though our method dramatically reduces the total number of random walks needed to solve PDEs relative to WoS(t), further investigation is needed to also reduce the large amounts of redundant computation performed across walks starting from the boundary---it is indeed wasteful to use each sphere in a walk to only estimate the solution at a single sample point at the very beginning of a walk.

\begin{figure*}
    \includegraphics[width=\textwidth]{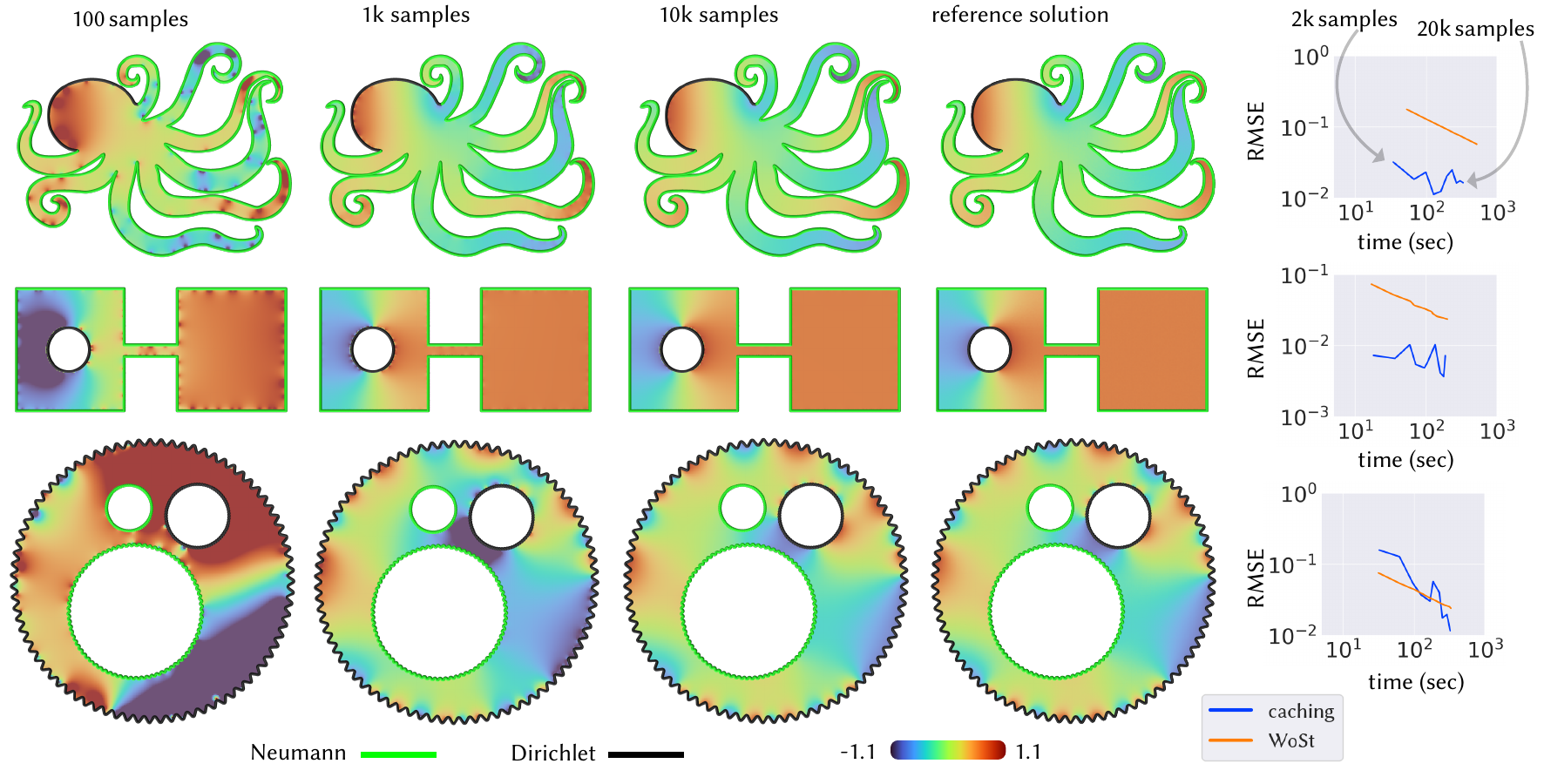}
    \caption{Our caching strategy amortizes the cost of long walks in Neumann dominated problems (\emph{top two rows}) and hence improves efficiency over point estimators like WoSt. However, in more Dirichlet dominated problems with higher frequency boundary conditions (\emph{bottom row}), efficiency drops due to shorter walk lengths, not importance sampling the singular functions in the BIE (\eqref{BoundaryIntegralEquation}), and noise from estimates of $\nicefrac{\partial u}{\partial n}$ on the boundary.}
    \label{fig:Gallery}
\end{figure*}

\begin{figure*}
    \includegraphics[width=\textwidth]{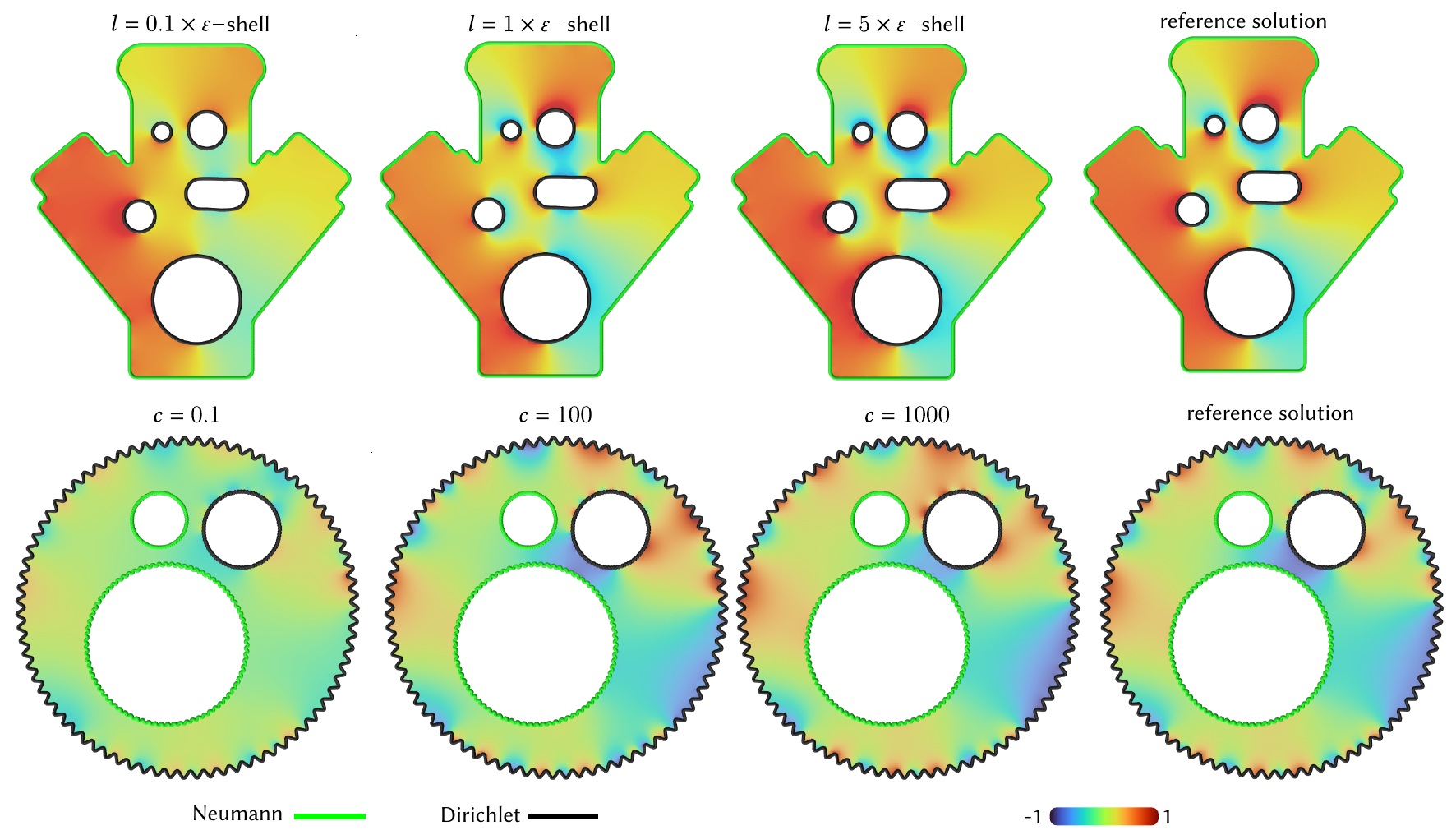}
    \caption{\emph{Top:} Setting the $l$ offset parameter to $0.1 \times \varepsilon$ effectively sets each Dirichlet boundary sample's $\nicefrac{\partial u}{\partial n}$ estimate to zero, as balls centered at each sample point are contained entirely inside the epislon shell---this biases the solution estimate inside the domain. Bias diminishes with increasing offset values. \emph{Bottom:} Smaller values of the bound $c$ for $\nicefrac{\partial G}{\partial n}$ in \eqref{PoissonKernelIntegralClamped} suppress singular artifacts near the boundary but biases interior estimates without our correction strategy, shown here on a model scaled to fit inside a unit sphere.}
    \label{fig:Ablations}
\end{figure*}

%\begin{tabular}{r|l}
%   Column width & \the\columnwidth \\
%   Text width & \the\textwidth \\
%   Font size & \makeatletter\f@size pt \\
%\end{tabular}

\begin{acks}
    This work was generously supported by nTopology and Disney Research, NSF awards 1943123, 2212290 and 2008123, Alfred P. Sloan Research Fellowship FG202013153, a Packard Fellowship, NSF Graduate Research Fellowship DGE2140739, and an NVIDIA Graduate Fellowship.
\end{acks}

% Bibliography
\bibliographystyle{ACM-Reference-Format}
\bibliography{BoundaryValueCaching}

\appendix

\section{Green's Functions and Their Derivatives}\label{app:GreensFns}
Here we provide the free-space Green's functions and their derivatives our caching strategy uses to solve Poisson and screened Poisson equations.

\subsection{Poisson Equation}
The free-space Green's functions in 2D and 3D for two points $x$ and $y$ are given by
\begin{equation}
    \label{eq:FreeSpaceGreensFnsPoisson}
    G_{2D}(x,y) = \frac{\log(r)}{2\pi}
    ,\qquad
    G_{3D}(x,y) = \frac{1}{4\pi r},
\end{equation}
where $r \equiv |y - x|$. Their gradients with respect to $x$ are
\begin{equation}
    \label{eq:FreeSpaceGreensFnsGradientsPoisson}
    \frac{\partial G_{2D}(x,y)}{\partial x} = \frac{x - y}{2\pi r^2},
    \qquad
    \frac{\partial G_{3D}(x,y)}{\partial x} = \frac{x - y}{4\pi r^3}.
\end{equation}
The normal derivatives with respect to $y$ are
\begin{equation}
    \label{eq:FreeSpacePoissonKernels}
    \frac{\partial G_{2D}(x,y)}{\partial n_y} = \frac{n_y \cdot \left(y - x\right)}{2\pi r^2},
    \qquad
    \frac{\partial G_{3D}(x,y)}{\partial n_y} = \frac{n_y \cdot \left(y - x\right)}{4\pi r^3},
\end{equation}
where $n_y$ is a unit normal direction associated with $y$. Expressions for $\nicefrac{\partial^2 G}{\partial x \partial n_y}$ are given by
\begin{align}
    \frac{\partial^2 G_{2D}(x,y)}{\partial x\ \partial n_y} &= 2 \frac{n_y \cdot \left(y - x\right)}{2\pi r^4}\left(y - x\right)\ -\ \frac{n_y}{2\pi r^2},\\
    \frac{\partial^2 G_{3D}(x,y)}{\partial x\ \partial n_y} &= 3 \frac{n_y \cdot \left(y - x\right)}{4\pi r^5}\left(y - x\right)\ -\ \frac{n_y}{4\pi r^3}.\label{eq:FreeSpaceGreensFns2ndDerivativePoisson}
\end{align}

\subsection{Screened Poisson Equation}
We denote by $K_n$ (for $n = 0, 1, \ldots$) the modified Bessel functions of the second kind. For a screened Poisson equation with a screening coefficient $\sigma \in \mathbb{R}_{\geq 0}$, the 2D and 3D free-space Green's functions are given by
\begin{equation}
    \label{eq:FreeSpaceGreensFnsScreenedPoisson}
    G_{2D}^{\sigma}(x,y) = \frac{K_0(r\sqrt{\sigma})}{2\pi},
    \qquad
    G_{3D}^{\sigma}(x,y) = \frac{e^{-r\sqrt{\sigma}}}{4\pi r}.
\end{equation}
Their gradients with respect to $x$ are
\begin{align}
    \frac{\partial G_{2D}^{\sigma}(x,y)}{\partial x} &= Q^{\sigma}_{2D}(x, y)\ \frac{\partial G_{2D}(x,y)}{\partial x},\\
    \frac{\partial G_{3D}^{\sigma}(x,y)}{\partial x} &= Q^{\sigma}_{3D}(x, y)\ \frac{\partial G_{3D}(x,y)}{\partial x},\label{eq:FreeSpaceGreensFnsGradientsScreenedPoisson}
\end{align}
where $\nicefrac{\partial G}{\partial x}$ are the corresponding free-space gradients for the Poisson equation, and
\begin{align}
    Q^{\sigma}_{2D}(x, y) &\equiv K_1(r\sqrt{\sigma})\ r\sqrt{\sigma},\\
    Q^{\sigma}_{3D}(x, y) &\equiv e^{-r\sqrt{\sigma}}\left(r\sqrt{\sigma} + 1\right).
\end{align}
The normal derivatives with respect to $y$ are
\begin{align}
    \frac{\partial G_{2D}^{\sigma}(x,y)}{\partial n_y} &= Q^{\sigma}_{2D}(x, y)\ \frac{\partial G_{2D}(x,y)}{\partial n_y},\\
    \frac{\partial G_{3D}^{\sigma}(x,y)}{\partial n_y} &= Q^{\sigma}_{3D}(x, y)\ \frac{\partial G_{3D}(x,y)}{\partial n_y}.\label{eq:FreeSpaceScreenedPoissonKernels}
\end{align}
Applying the product rule to the expressions above yields $\nicefrac{\partial^2 G^{\sigma}}{\partial x \partial n_y}$.

\section{Open Domains and Double-Sided Boundary Conditions}\label{app:BIEDoubleSided}
For a domain $\Omega \subset \mathbb{R}^N$ with open boundaries and double-sided boundary conditions, the BIE in \eqref{BoundaryIntegralEquation} for can be generalized to \citep{Costabel:1987:BEM}
\begin{align}
    u(x) &= \int_{\partial \Omega} \frac{\partial G}{\partial n^+}(x, z)\ \left[u^+(z) - u^-(z)\right]\nonumber\\
         &- G(x, z)\ \left[\frac{\partial u^+}{\partial n^+}(z) - \frac{\partial u^-}{\partial n^-}(z)\right]\ dz\nonumber\\
         &+ \int_{\Omega} G(x, y)\ f(y)\ dy,\label{eq:BoundaryIntegralEquationDoubleSided}
\end{align}
for any point $x \in \Omega$, where $n^+$ and $n^-$ denote unit outward and inward facing normals on $\partial \Omega$ respectively, and $u^+$ and $u^-$ represent corresponding solution values on either side of $\partial \Omega$. From a sample reuse perspective, we generate uniformly distributed source samples inside $\Omega$ as usual (if $\Omega$ is unbounded, we create a bounding box around it), but now use two separate sets of boundary samples associated with the two normal directions $n^+$ and $n^-$ on $\partial \Omega$ respectively. As discussed in \citet[Appendix B]{Sawhney:2023:WalkOnStars}, we can use WoSt to estimate $u^+$ and $u^-$ by launching random walks on either side of the Neumann boundary $\partial \Omega_N$. Similar to \secref{BoundarySpecification}, we generate boundary samples on two offset Dirichlet boundaries $\partial \Omega_D^l$ and $\partial \Omega_D^{-l}$ to estimate $\partial u^+/\partial n^+$ and $\partial u^-/\partial n^-$ with WoS respectively. With this setup, we can then compute a solution estimate for any evaluation point $x \in \Omega$ by using all source and boundary samples, except for points that are within a distance $l$ to $\partial \Omega_D$ where we use WoSt to compute pointwise solution estimates.

\section{Mitigating Singular Artifacts from the Green's Function}\label{app:GreensFnArtifacts}
We follow a similar recipe to \eqref{PoissonKernelIntegralClamped} to mitigate localized artifacts that arise from not importance sampling the Green's function in the BIE's source integral. In particular, if $c$ is a positive user-specified bound and $G_{c} \equiv \min(-c, \max(c, G))$, then we can rewrite the source integral over a region $R$ as follows:
\begin{equation}
    \label{eq:GreensFnIntegralClamped}
    \int_{R} G^B_{c}(x, y)\ f(y)\ dy\ +\ \int_{R} \left[G^B(x, y) - G^B_{c}(x, y)\right] f(y)\ dy,
\end{equation}
where rather than the free-space Green's function $G$, we now use the Green's function of a ball $G^B$ (see \citet[Appendix A]{Sawhney:2023:WalkOnStars}). The reason we use $G^B$ instead of $G$ is because it can be importance-sampled for $y$ \citep[Section 1.4, Supplemental]{Sawhney:2022:VCWoS}. It also shares the same normal derivative as its free-space counterpart, which allows $G^B$ to be used in the BIE's boundary integral as well. We choose the ball $B$ to be centered at the evaluation point $x$ such that it contains the region $R$ inside it. The first integral in \eqref{GreensFnIntegralClamped} is then estimated as usual by using uniformly distributed source samples in $R$. The second integral is estimated by sampling $G^B$ for $y$ and setting $f = 0$ if $y \notin R$.

\end{document}